\documentclass[runningheads]{llncs}
\usepackage{graphicx}

\usepackage{cite}
\usepackage{amsmath,amssymb,amsfonts}
\usepackage{algorithmic}
\usepackage{textcomp}
\usepackage{xcolor}

\usepackage[T1]{fontenc}
\usepackage[utf8]{inputenc}
\usepackage[english]{babel}
\usepackage{chemformula}
\usepackage{url}
\usepackage{pifont}
\usepackage{amsmath}
\usepackage[output-decimal-marker={.}]{siunitx}
\usepackage[hidelinks]{hyperref}
\usepackage{microtype}
\usepackage{soul}
\usepackage{braket}
\usepackage{xcolor}
\usepackage{booktabs}
\usepackage{float}
\usepackage{subcaption}
\usepackage{bm}
\usepackage{lscape}
\usepackage{threeparttable}
\usepackage{comment}
\usepackage{csquotes}
\begin{document}
\title{Ab initio Calculation of Binding Energies of Interstellar Sulphur-Containing Species on Crystalline Water Ice Models}
\titlerunning{Binding Energies of Interstellar Sulphur-Containing Species}
%
\author{Jessica Perrero\inst{1}\orcidID{0000-0003-2161-9120} \and
Albert Rimola\inst{1}\orcidID{0000-0002-9637-4554} \and Marta Corno\inst{2}\orcidID{0000-0001-7248-2705} \and
Piero Ugliengo\inst{2}\orcidID{0000-0001-8886-9832}  }
\authorrunning{J. Perrero et al.}
%
\institute{Departament de Química, Universitat Autònoma de Barcelona, E-08193 Bellaterra, Catalonia, Spain \\
\email{jessica.perrero@uab.cat \\ albert.rimola@uab.cat}\\
 \and
Dipartimento di Chimica, Università degli Studi di Torino,
via P. Giuria 7, I-10125, Torino, Italy \\
\email{marta.corno@unito.it \\ piero.ugliengo@unito.it}}
\maketitle              
\begin{abstract}
There are different environments in the interstellar medium (ISM), depending on the density, temperature and chemical composition. Among them, molecular clouds, often referred to as the cradle of stars, are paradigmatic environments relative to the chemical diversity and complexity in space. Indeed, there, radio to far-infrared observations revealed the presence of several molecules in the gas phase, while near-infrared spectroscopy detected the existence of submicron sized dust grains covered by \ch{H2O}-dominated ice mantles. The interaction between gas-phase species and the surfaces of water ices is measured by the binding energy (\emph{BE}), a crucial parameter in astrochemical modelling. In this work, the \emph{BEs} of a set of sulphur-containing species on water ice mantles have been computed by adopting a periodic \textit{ab initio} approach using a crystalline surface model. The Density Functional Theory (DFT)-based B3LYP-D3(BJ) functional was used for the prediction of the structures and energetics. DFT \emph{BEs} were  refined by adopting an ONIOM-like procedure to estimate them at CCSD(T) level toward complete basis set extrapolation, in which a very good correlation between values has been found. Moreover, we show that geometry optimization with the computationally cheaper HF-3c method followed by single point energy calculations at DFT to compute the \emph{BEs} is a suitable cost-effective recipe to arrive at \emph{BE} values of the same quality as those computed at full DFT level. Finally, computed data were compared with the available literature data. 
\end{abstract}

\section{Introduction}
The interstellar medium (ISM) is the region between stars constituted by matter and radiation. The interstellar matter consists of either gaseous (atoms and molecules) or solid (dust grains of silicate or carbonaceous material) components. Several environments can be defined in the ISM according to their physical conditions, amongst them the dense molecular clouds, cold (10 K) and rarefied (\SI{e3}{\per\cubic\centi\meter} - \SI{e7}{\per\cubic\centi\meter}) regions in which stars form. Because of the very low temperature, species heavier than He disappear from the gas phase due to freeze-out (adsorption) on dust grain surfaces, forming a thick layer of ices. The ice composition is dominated by \ch{H2O} ice, but also by small molecules like CO, \ch{CO2}, \ch{CH3OH}, \ch{CH4} and \ch{NH3} \cite{boogert:2015}. The significant presence of hydrogenated species on the ice mantles is due to the hydrogenation reactions taking place on the ice surfaces and involve atoms and small molecules that are adsorbed on grains \cite{caselli:2012}. 

Dust grains play a fundamental role in the chemistry of the ISM. Beyond their physical properties, like UV absorption and radio emission, they can help the occurrence of chemical reactions in three ways: i) they can concentrate chemical species on the surfaces, facilitating the encountering of the reactive species; ii) they can act as chemical catalysts, decreasing the activation energy of chemical reactions; and iii) they can be a third body, allowing the dissipation of the energy released by exothermic reactions without undermining the stability of the products. 

Molecules on the grain surfaces require some energy to desorb. This energy can be thermal, chemical or radiative, but it has to exceed the interaction energy between the molecule and the surface since the evaporation rate is proportional to $exp[-E_B/(k~ T_{dust})]$, where $T_{dust}$ < 10 K and the binding energies are typically $E_B$ > 1000 K. If the amount of energy is a fraction of the binding energy, species can diffuse on the surfaces, which is one of the prerequisite for their subsequent reactivity \cite{das:2018}. Accordingly, one of the most important parameters in the description of the ISM are the binding energies (\emph{BEs}), i.e., the interaction energy of chemical species on dust grains . Molecules adsorbed on grains can interact with them in two main ways: through rupture/formation of chemical bonds, so-called chemisorption, or through weak interactions, resulting in  physisorption. This latter mechanism is the focus of this work and what our \emph{BEs} are referred to.
 \emph{BE} governs whether molecules remain stick on the grain surfaces as well as whether they diffuse and, therefore, react with other species. Thus, \emph{BEs} hugely affect the gas and ices composition. Moreover, they are essential parameters to run astrochemical models at large aimed to understand the chemical composition of the ISM. \\

The importance of a careful estimation of \emph{BEs} is thus evident. However, they have usually been estimated simply as a sum of the \emph{BEs} of the single atoms or functional groups forming the molecule (in turn estimated from their polarizability when interacting with bare grain surfaces). In addition, the experimental procedure to collect \emph{BEs} is not completely trustworthy, considering the impossibility to reproduce properly the ISM conditions in a laboratory chamber \cite{das:2018,penteado:2017}. Ferrero et al. (2020) \cite{ferrero:2020}, have used quantum chemical computations as an alternative way to obtain accurate \emph{BE} values for a set molecular species on water ices. The aim of this work is to enlarge the network of \emph{BEs}, focusing on eight sulphur-containing molecules that play a major role in the dense clouds.

\section{Methods}\label{sec:methods}

We adopted a periodic approach to model the adsorption of the sulphur-containing molecules (all of them being closed-shell species) on the ice surface. The hybrid B3LYP DFT functional \cite{lyp,becke:1988,becke:1993}, with the addition of Grimme's D3 empirical correction with Becke-Johnson (BJ) damping scheme to account for the dispersion interactions \cite{sure:2013,grimme:2010,grimme:2011}, in combination with the Ahlrichs triple zeta quality VTZ basis set, supplemented with a double set of polarization functions (Ahlrichs-VTZ*), was used for geometry optimizations and \emph{BE} calculations (hereafter referred to as full B3LYP-D3(BJ) theory level).

Because of the high computational cost of DFT calculations, we also adopted the computationally cheaper semi-empirical HF-3c method in two ways \cite{sure:2013}. This method is based on the Hartree-Fock (HF) energy computed with a minimal basis set, to which three empirical corrections (3c) are added.
In the first case, HF-3c was used both in the optimization and the \emph{BE} calculation (HF-3c//HF-3c theory level). In the second one, DFT single point energy calculations on the HF-3c optimized geometries were performed to compute \emph{BEs} (B3LYP-D3(BJ)//HF-3c theory level). The latter approach was used to validate its performance in prevision of applying the same approach in amorphous ice models, which are prohibitively too expensive to be simulated at full DFT level.

Finally, to prove that DFT \emph{BEs} are accurate enough, the single- and double- electronic excitation coupled-cluster method with an added perturbative description of triple excitations (CCSD(T)), in combination with a correlation consistent basis set extrapolated to complete basis set (CBS) limit, was used by applying a local correction at this level adopting an ONIOM2-correction approach \cite{dapprich:1999}, in which a small part of the system (i.e., the adsorbed molecule plus a few interacting water molecules) was treated at CCSD(T), while the rest at B3LYP-D3(BJ).

The \emph{BE} is defined as the opposite of the interaction energy $\Delta$E, which were corrected for the basis set superposition error (BSSE), due to using a finite basis set of localized gaussian functions. 
\begin{equation}
BE = -\Delta E^{CP}
\end{equation}
\begin{equation}
    \Delta E^{CP} = \Delta E - BSSE
\end{equation}

All periodic calculations were performed with the \textsc{crystal17} code \cite{crystal}, while CCSD(T) calculations were performed with the \textsc{gaussian16} code \cite{gaussian}.

\section{Results}

Interstellar ices are thought to be mostly formed by amorphous solid water \cite{boogert:2015}. We, however, have chosen a crystalline model for two main reasons: (i) crystalline structure are well defined due to symmetry constraints and computationally cheap, and (ii) no definite structure of an amorphous ice model is available. Nevertheless, regions rich in crystalline ices have been observed in protoplanetary disks and stellar outflows \cite{molinari:1999,terada:2012}. Our ice surface model derives from the bulk of the proton-ordered P-ice, which was cut along the (010) surface defining a 2D periodic slab model \cite{casassa:1997,zamirri:2018}. The thickness of the ice (\SI{10.9}{\angstrom}) was chosen to converge its surface energy. The (010) P-ice supercell slab model consists of twelve atomic layers (24 water molecules) and its cell parameters are |a| = \SI{8.980}{\angstrom} and |b| = \SI{7.081}{\angstrom}. The structure of the ice is such as to ensure a null electric dipole along the non-periodic z-axis. This is a direct consequence of the symmetry of the system, which shows two identical faces both at the top and at the bottom of the model. Therefore, the adsorption was modeled only onto one face of the system.

\subsection{BEs of the S-containing species}
\begin{figure}[htp]
    \centering
    \begin{subfigure}[t]{0.99\linewidth}
    \includegraphics[width=0.9\linewidth]{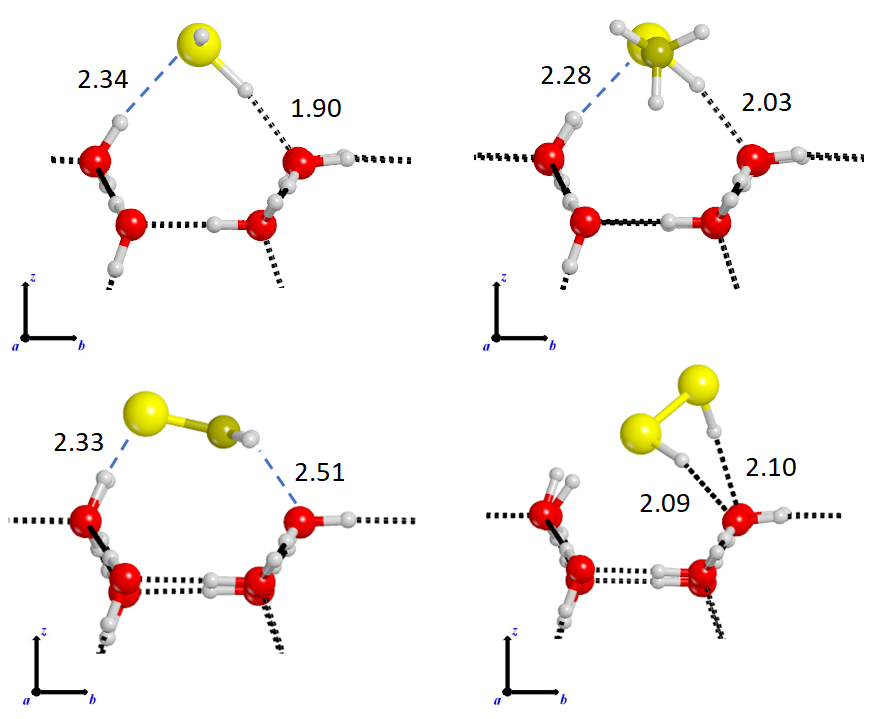}
    \caption{Side view of B3LYP-D3(BJ)/Ahlrichs-VTZ* optimized geometries of \ch{H2S}, \ch{CH3SH}, \ch{H2CS} and \ch{H2S2}.
    }
    \label{gruppo1}
    \end{subfigure}
    \begin{subfigure}[b]{0.99\linewidth}
    \includegraphics[width=0.9\linewidth]{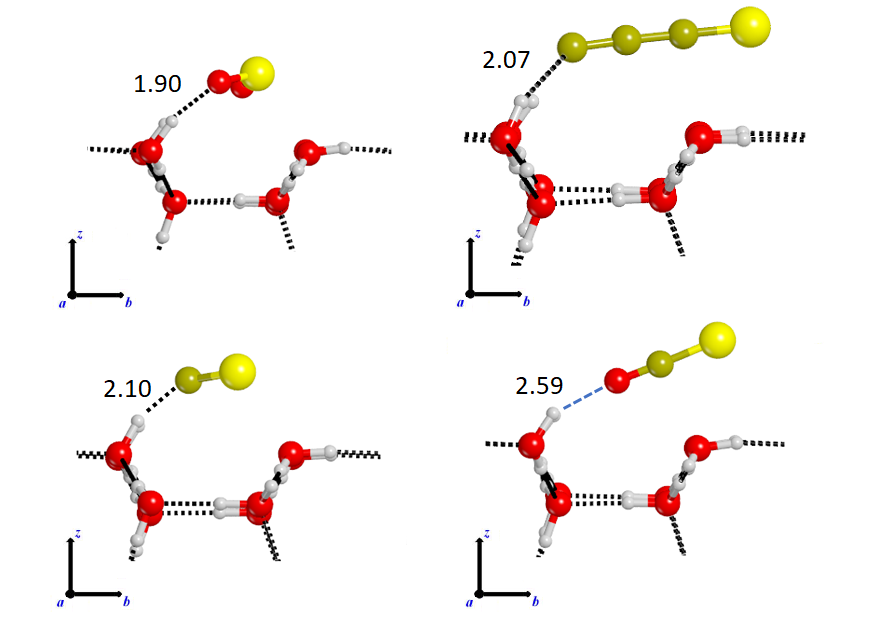}
    \caption{Side view of B3LYP-D3(BJ)/Ahlrichs-VTZ* optimized geometries of \ch{SO2}, \ch{C3S}, \ch{CS} and \ch{OCS}.
    }
    \label{gruppo2}
    \end{subfigure}
    \caption{Optimized geometries of the 8 S-bearing species adsorbed on the ice crystalline slab (010). Hydrogen bond distances are given in Angstrom. Color legend: red, O; yellow, S; white, H; brown, C.}
\end{figure}

Here the \emph{BEs} of 8 sulphur-containing species, i.e., \ch{H2S}, \ch{CH3SH}, \ch{H2S2}, \ch{H2CS}, \ch{C3S}, \ch{SO2}, \ch{CS} and \ch{OCS}, on the crystalline (010) water ice surface model were computed. Figures \ref{gruppo1} and \ref{gruppo2} show the optimized geometries of the computed complexes at full B3LYP-D3(BJ) level, while Table \ref{tab:BE} reports the computed \emph{BE} for these species.

According to the adsorption geometries, these species can be divided in two groups: i) molecules that can act as both hydrogen bond (H-bond) acceptors and donors, given the presence of an electronegative sulphur atom and one or more hydrogen atoms (\ch{H2S}, \ch{CH3SH}, \ch{H2S2} and \ch{H2CS}); and ii) molecules that can only act as H-bond acceptors, through an electronegative atom which is never the sulphur one (\ch{C3S}, \ch{SO2}, \ch{CS} and \ch{OCS}). 
We would like to highlight that this classification does not always correspond with the electrostatic (non dispersive) contribution to the \emph{BEs}. Indeed, not only  H-bond contributes to the definition of the electrostatic energy, but also the interaction between permanent dipoles and higher order multipoles. Additionally, dispersive forces are stronger as the volume of the molecule and the interaction surface with the ice increase, resulting in the articulated panorama appearing in Table \ref{tab:BE}. While the molecules both accepting and donating an H-bond show a \emph{BE} spanning the limited range between 40 and 52 kJ mol$^{-1}$, the other group presents more variability. In fact, not only these molecules show very different dipoles, but they also differentiate in their mass and shape, giving rise to a dispersive force that can account for between 50\% and 100\% of the \textit{BE}.
The case of OCS is of particular interest because, by having the smallest dipole moment in the set of molecules, it is the less bound species to the ice. Not only it has the smallest binding energy, but also its interaction with the ice is almost completely (>90\%) due to the dispersive forces.\\

\begin{table}[htp]
    \centering
        \caption{B3LYP-D3(BJ)/Ahlrichs-VTZ* computed \textit{BEs} (in kJ mol$^{-1}$) of the sulphur-containing species on the P-ice (010) (2x1) super cell model. The dispersive and non-dispersive contribution to the \textit{BE} are also listed. Whenever more than one adsorption geometry was found, all \textit{BEs} are listed and separated by a slash.}
    \begin{tabular}{l@{\hspace{1.2cm}} c@{\hspace{1.2cm}} c@{\hspace{1.2cm}} c}
    \toprule
    Molecule & Total \textit{BE} &  No dispersion & Only dispersion(\%)\\
    \midrule
    \ch{H2S} & 43.8  & 27.7  & 16.1(37\%)\\
    \ch{CH3SH}   & 51.4 & 24.6  & 26.8(52\%) \\
    \ch{H2CS} &  40.3 / 51.9 &  18.5 / 23.7 & 21.8(54\%) / 28.2(54\%)\\
    \ch{H2S2} & 45.2 / 51.2 & 13.5 / 16.1  & 31.7(70\%) / 35.1(69\%) \\
    \midrule
      \ch{SO2} &  57.2 & 30.0  & 27.2(48\%) \\
     \ch{CS} & 38.1 & 13.3  & 24.8(65\%) \\
     \ch{C3S} & 50.2 / 55.9& 13.0 / 17.7  & 37.2(74\%) / 38.2(68\%)\\
    \ch{OCS} & 25.0 / 28.6 & -1.8 / 2.8  & 26.8(107\%) / 25.8(90\%)  \\
    \bottomrule
    \end{tabular}
    \label{tab:BE}
\end{table}

\begin{figure}[htp]
    \centering
    
\end{figure}

\subsection{HF-3c method}
As mentioned above, interstellar ice grains are mostly considered as amorphous solid water (ASW) systems. Unfortunately, ASW can only be simulated by requesting unit cells large enough to provide sufficient structural variability to mimic the amorphous nature. This increases the request of computer resources, making DFT methods almost impractical. Therefore, here we check our DFT results against those obtained with HF-3c, a much cheaper methodology to potentially be used when dealing with amorphous systems. \\

\begin{figure}[h!]
    \centering
    \begin{subfigure}[t]{0.9\linewidth}
    \includegraphics[width=1.0\linewidth]{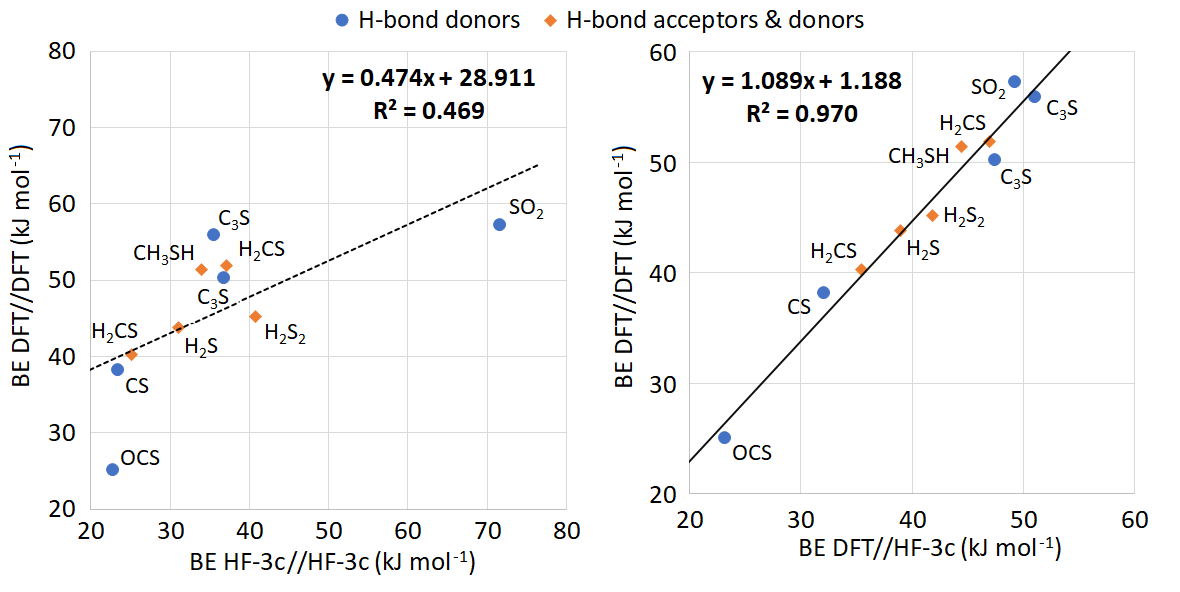}
    \caption{Left panel: correlation between \emph{BEs} computed at HF-3c//HF-3c level versus B3LYP-D3(BJ)//B3LYP-D3(BJ). Right panel: correlation between \emph{BEs} computed at B3LYP-D3(BJ)//HF-3c level versus B3LYP-D3(BJ)//B3LYP-D3(BJ). \emph{BEs} are in kJ mol$^{-1}$.}
    \label{hf3c}
    \end{subfigure}
    \begin{subfigure}[t]{0.8\linewidth}
    \centering
    \includegraphics[width=0.5\linewidth]{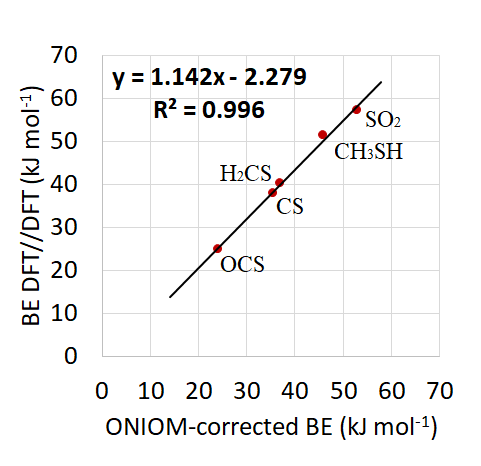}
    \caption{Correlation between ONIOM-corrected \textit{BEs} versus B3LYP-D3(BJ)//B3LYP-D3(BJ) computed energies. \textit{BEs} are in kJ mol$^{-1}$. The four species were chosen in order to span the entire range of \textit{BEs} computed in this work and in order to select two species form each group.}
    \label{ccsdt}
    \end{subfigure}
    \caption{Correlation of DFT//DFT \textit{BEs} versus two approaches involving HF-3c (panel a) and versus ONIOM-corrected \textit{BEs} (panel b).}
\end{figure}

From the optimized DFT structures, we proceeded through the two HF-3c ways explained in the Methods section, i.e., at HF-3c//HF-3c and at B3LYP-D3(BJ)//HF-3c theory levels. The obtained results are summarized in Table \ref{tab:HF-3c}. Figure \ref{hf3c} (left panel) shows the comparison between the \textit{BE} computed at full B3LYP-D3(BJ)  with the results computed at HF-3c//HF-3c. The correlation is only coarse, as it would be expected from the various approximations included in the HF-3c definition. In contrast, the performance of the B3LYP-D3(BJ)//HF-3c method is very good (see Figure \ref{hf3c}, right panel), showing its applicability to compute \emph{BEs} on amorphous surface models in terms of computational cost and quality of the results, allowing moreover a wide exploration of different binding sites in an affordable way.

\begin{table}[htp]
    \centering
        \caption{Summary of the computed \textit{BEs} (in kJ mol$^{-1}$) of the set of chosen species using the P-ice (010) (2x1) supercell.}
    \begin{tabular}{l c@{\hspace{1.0cm}} c@{\hspace{1.0cm}} c}
    \toprule
    Molecule & DFT//DFT &  HF-3c//HF-3c & DFT//HF-3c\\
    \midrule
    \ch{H2S} & 43.8  & 31.0  & 38.9 \\
    \ch{CH3SH}   & 51.4 &  34.0 & 44.4  \\
    \ch{H2CS} &  40.3 / 51.9 &  25.1 / 37.2 & 35.4 / 46.9\\
    \ch{H2S2} & 45.2 / 51.2 &  40.8 & 41.8 \\
    \midrule
     \ch{SO2} &  57.2 & 71.6  &  49.3\\
     \ch{CS} & 38.1 & 23.6  & 32.1 \\
     \ch{C3S} & 50.2 / 55.9& 36.9 / 35.6  & 47.1 51.1 \\
    \ch{OCS} & 25.0 / 28.6 & 22.9   & 23.2  \\
    \bottomrule
    \end{tabular}
    \label{tab:HF-3c}
\end{table}

\subsection{CCSD(T) correction}
We finally performed some calculations at CCSD(T)/CBS (ONIOM2-correction, see Section \ref{sec:methods}) to check the accuracy of the B3LYP-D3(BJ) functional in describing the local interactions between the species and the ice and, at the same time, to refine the computed \emph{BEs}. We chose to compute the correction only for some species, with the aim to span as much as possible the range of \textit{BEs} and to sample geometries due to different contributions of dispersion forces. We are aware of the fact that five points are few for a meaningful statistics. However, when plotting the ONIOM2-corrected CCSD(T)/CBS \emph{BEs} against those computed at the full B3LYP-D3(BJ) level, one can see an almost perfect correlation between the values, regardless of the nature of their interaction (see Figure \ref{ccsdt}). Therefore, our results evidence the accuracy and reliability of the full B3LYP-D3(BJ) procedure adopted and, by extension, of the B3LYP-D3(BJ)//HF-3c scheme.

\section{Discussion}

\begin{figure}[hbt]
    \centering
    \includegraphics[width=0.75\linewidth]{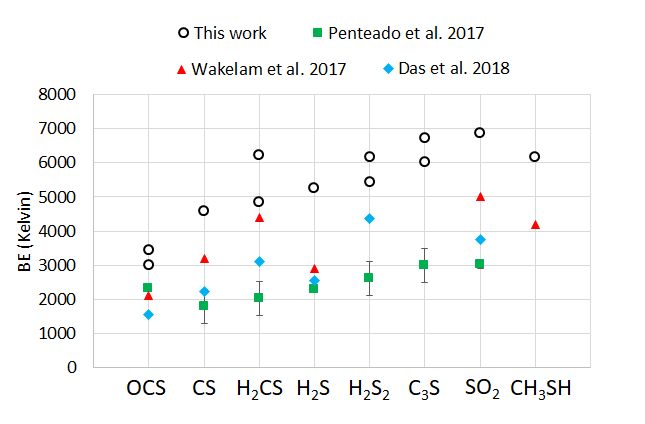}
    \caption{\textit{BEs} of the 8 S-bearing species (in Kelvin) at B3LYP-D3(BJ)//B3LYP-D3(BJ) level compared with those published by Wakelam et al. (2017) (red triangles), Das et al. (2018) (blue rhombus) and Penteado et al. (2017) (green squares, with error bars). \cite{penteado:2017,das:2018,wakelam:2017}}
    \label{literature}
\end{figure}

Our computed \emph{BEs} have been compared with the available literature data in Figure \ref{literature}. In 2017, Penteado et al. \cite{penteado:2017} presented a list of recommended \emph{BEs}, collecting data from previous works and adding an uncertainty range to each value. For \ch{H2S}, OCS and \ch{SO2}, the \emph{BEs} were estimated from the experimental work of Collings et al. (2004) \cite{collings:2004}, while for all other species, data were based on the work of Hasegawa \& Herbst (1993) \cite{hasegawa:1993}, which collected \emph{BEs} from previous works, including Allen \& Robinson 1977 \cite{allen:1977}. The uncertainty range for these values was set to half the \emph{BE} value for species whose \emph{BE} is less than 1000 K and to 500 K for all other cases. Comparison between these \emph{BE} estimates with our computed \emph{BEs} shows discrepancies, our values being systematically larger. This inconsistency should be taken with care, given the origin of the data published by Penteado et al. (2017). The cases of \ch{H2S}, \ch{OCS} and \ch{SO2},  the ones extrapolated from thermal desorption process (TPD) experiments by Collings et al. (2004), are of interest. In the experiments, the species were deposited onto a pre-existing film of \ch{H2O} at 8-10 K with an exposure that ensured a multilayer growth of each species. This is at variance with our assumption of a submonolayer adsorption. Therefore, while for OCS the difference amounts to about 30\%, for \ch{H2S} and \ch{SO2} this percentage goes up to almost 60\%. An additional reason explaining these large differences can be tracked down to the crystalline nature of our computer model compared to experimental data measured for AWS ice. The paucity of different adsorption sites at the crystal ice and the enhanced hydrogen bond cooperativity compared with amorphous ices give \emph{BEs} values higher than any experimental estimate. Additionally, the fact that our \emph{BE} values do not account for zero-point energy (ZPE) corrections contributes to enlarge the discrepancies. The comparison with values based on the additive principle by Allen and Robinson in 1997 \cite{allen:1977} present an average difference of 50\%. Given the huge difference in the theoretical background standing behind these values, it is not straightforward explaining the differences. However, we can advance that the additive principle employed by Allen and Robinson \cite{allen:1977} to estimate the \textit{BEs}, although being a reasonable approach in 1977, it is currently not suitable to obtain \textit{BEs} since the interaction of a molecule with a surface cannot be described as the sum of the interactions of the single atoms constituting the molecule.

In 2017, Wakelam et al. proposed a new chemical model to compute the binding energies \cite{wakelam:2017}. Given the difficulty to reproduce a reliable structure of an amorphous ice, it was assumed that \emph{BEs} on ASW are proportional to the interaction energy with one single water molecule, without ZPE nor BSSE correction. The approximate nature of this model was recovered by correcting the computed \emph{BEs} through a scaling factor from the best fit with selected experimental \emph{BE} values. The general trend by comparing our computed \emph{BE} values with those of Wakelam et al. (2017) is a percentage difference of about 30\%, the our values being larger. Although being corrected, it seems clear that using one water molecule to model an entire water ice surface is not a suitable choice since different important aspects such as H-bond cooperativity and surface long-range effect are neglected. 

In 2018, Das et al. \cite{das:2018} simulated the water surface using four, and in some cases, six water molecules. For S-bearing species, only \ch{H2S} and \ch{OCS} were considered at MP2/aug-cc-pVDZ level on the water tetramer. Comparison of these values with our results with shows large differences. Reasons are twofold: i) the P-ice (010) surface exploits full hydrogen bond cooperativity, largely missing in the water tetramer; and ii) the adopted quantum mechanical methods are significantly different. Moreover \emph{BEs} by Das et al. (2018) were not corrected for BSSE, which sure contribute to these discrepancies, nor for the ZPE \cite{das:2018}.  

The work of Ferrero et al. (2020) \cite{ferrero:2020}, presented for a set of chemical species a \emph{BE} distribution by modeling their adsorption on both a crystalline and an amorphous surface ice models. Results showed  that those on the crystalline ice were usually in the upper region of the range. Thus, we expect a similar behaviour for the adsorption of S-bearing species. Therefore, literature values should then be included in the range spanned by the distributions. With that being said, it is clear that the computation of \emph{BEs} on an amorphous water ice model is necessary and already scheduled for the future, especially because these values have a pivotal role in the accuracy of the predictions provided by astrochemical models that deal with the chemical evolution in the ISM.

\section{Conclusions}
In this work we computed the \textit{BEs} of 8 astrochemically-relevant S-containing species (i.e., \ch{H2S}, \ch{CH3SH}, \ch{H2S2}, \ch{H2CS}, \ch{C3S}, \ch{SO2}, \ch{CS} and \ch{OCS}) on water ice surfaces by means of quantum chemical calculations. We followed the same approach proposed by Ferrero et al. (2020) \cite{ferrero:2020}, in which the interstellar water ice was simulated through the (010) surface of the crystalline P-ice \cite{casassa:1997} as first step to subsequent studies in which amorphous ice systems will be employed. 
It is found that the S-containing species can be categorized in two groups: those acting as both H-bond acceptors and donors (\ch{H2S}, \ch{CH3SH}, \ch{H2S2}, \ch{H2CS}), and those acting only as H-bond acceptors (\ch{C3S}, \ch{SO2}, \ch{CS} and \ch{OCS}), in which the S atoms dos not participate in the interaction. Although \textit{a priori} one could think that the former group would present larger \emph{BEs} than the later ones, B3LYP-D3(BJ) results indicate that this is not the case since dispersive interactions are actually important, specially in the second group. CCSD(T) refinements on the \textit{BE} values adopting an ONIOM2-correction approach indicate that those obtained at full B3LYP-D3(BJ) level are accurate enough. Similarly, comparison between  B3LYP-D3(BJ)//HF-3c and full B3LYP-D3(BJ) \emph{BE} values indicate that the former method is an optimum cost-effective method to compute \emph{BEs}, in that case to be applicable in larger systems like the amorphous ones.

When comparing the computed \textit{BEs} with experimental and other theoretical values from the literature we found major discrepancies. They arise from: i) the adoption of a crystalline ice model, which emphasizes the H-bond cooperativity compared with the amorphous ones, and ii) the different accounting for of long range order effects, highly present in crystalline systems while lost in the random network of the amorphous ice. Furthermore, the paucity of adsorption sites due to symmetry constraints as provided by the crystalline systems gives \emph{BE} values that are at the upper extreme of the ranges compared with the larger surface morphological variability present in the amorphous ices. Such a constraint does not allow us to establish a statistical distribution of the \emph{BEs} values, which is compulsory if a fair comparison with the experimental TPD data is foreseen. It is clear that the uncertainties on these values require more studies on the subject, especially because the binding energies have a pivotal role in the description of the chemistry occurring in the ISM.

\section*{Acknowledgment}
This project has received funding from the European Research Council (ERC) under the European Union’s Horizon 2020 research and innovation programme (grant agreement No. 865657) for the project “Quantum Chemistry on Interstellar Grains” (QUANTUMGRAIN). MINECO (project CTQ2017-89132-P) DIUE (project 2017SGR1323) are acknowledged for financial support. This work is related to the project “Astro-Chemical Origins” (ACO) associated with the European Union’s Horizon 2020 research and innovation programme under the Marie Sklodowska-Curie grant agreement No. 811312. A. R. is indebted to the “Ramón y Cajal” program. The authors wish to thank the anonymous reviewers for their valuable suggestions.

\bibliographystyle{splncs04}
\bibliography{ch_biblio.bib}

\end{document}